\documentclass[lettersize,journal]{IEEEtran}
\usepackage{amsmath,amsfonts}
\usepackage{algorithmic}
\usepackage{algorithm}
\usepackage{array}
\usepackage[caption=false,font=normalsize,labelfont=sf,textfont=sf]{subfig}
\usepackage{textcomp}
\usepackage{stfloats}
\usepackage{url}
\usepackage{enumerate}
\usepackage{verbatim}
\usepackage{graphicx}
\usepackage{xcolor}
\usepackage{cite}
\usepackage{tabularx,booktabs,textcomp}
\usepackage{amsthm}

\usepackage{amsmath,amssymb,amsthm}

\begin{document}

\title{Planning of Fast Charging Infrastructure for Electric Vehicles in a Distribution System and Prediction of Dynamic Price}

\author{{K. Victor Sam Moses Babu$^{a}$, Pratyush Chakraborty$^{b}$, Mayukha Pal$^{c}$}
 %~\IEEEmembership{Member,~IEEE}
 %~\IEEEmembership{Senior Member,~IEEE}
        % <-this % stops a space
\thanks{(Corresponding author: $^{c}$Mayukha Pal)}
\thanks{$^{a}$K. Victor Sam Moses Babu is a Data Science Research Intern at ABB Ability Innovation Center, Hyderabad 500084, India and also a Ph.D. Research Scholar at the Department of Electrical and Electronics Engineering, BITS Pilani Hyderabad Campus, Hyderabad 500078, IN, (e-mail: victor.babu@in.abb.com).}
\thanks{$^{b}$Dr. Pratyush Chakraborty is an Asst. Professor in the Department of Electrical and Electronics Engineering, BITS Pilani Hyderabad Campus, Hyderabad 500078, IN, (e-mail:pchakraborty@hyderabad.bits-pilani.ac.in).}
\thanks{$^{c}$Dr. Mayukha Pal is with ABB Ability Innovation Center, Hyderabad-500084, IN, working as Global R\&D Leader – Cloud \& Analytics (e-mail: mayukha.pal@in.abb.com).}}

%\markboth{Journal of \LaTeX\ Class Files,~Vol.~, No.~, }%

\maketitle

\begin{abstract} The increasing number of electric vehicles (EVs) has led to the need for installing public electric vehicle charging stations (EVCS) to facilitate ease of use and to support users who do not have the option of residential charging. The public electric vehicle charging infrastructures (EVCIs) must be equipped with a good number of EVCSs, with fast charging capability, to accommodate the EV traffic demand, which would otherwise lead to congestion at the charging stations. The location of these fast-charging infrastructures significantly impacts the distribution system (DS). We propose the optimal placement of fast-charging EVCIs at different locations in the distribution system, using multi-objective particle swarm optimization (MOPSO), so that the power loss and voltage deviations are kept at a minimum. Time-series analysis of the DS and EV load variations are performed using MATLAB and OpenDSS. We further analyze the cost benefits of the EVCIs under real-time pricing conditions and employ an autoregressive integrated moving average (ARIMA) model to predict the dynamic price. The simulated test system without any EVCI has a power loss of 164.36 kW and squared voltage deviations of 0.0235 p.u. Using the proposed method, the results obtained validate the optimal location of 5 EVCIs (each having 20 EVCSs with a 50kWh charger rating) resulting in a minimum power loss of 201.40 kW and squared voltage deviations of 0.0182 p.u. in the system. Significant cost benefits for the EVCIs are also achieved, and an R-squared value of dynamic price predictions of 0.9999 is obtained. This would allow the charging station operator to make promotional offers for maximizing utilization and increasing profits. 
\end{abstract}

\begin{IEEEkeywords}
ARIMA, electric vehicle, fast charging infrastructure, multi-objective optimization, particle swarm optimization.
\end{IEEEkeywords}

\section{Introduction}
\label{section:Introduction}

\subsection{Background and Motivation}
The popularity of electric vehicles (EVs) has been growing in recent years. The main driving factor is their environmental benefits, as they produce zero emissions \cite{EV_purchase}. EVs are widely recognized as a promising solution for reducing air pollution and greenhouse gas emissions from the transportation sector. In addition, the falling cost of batteries and the improving performance of EVs make them more attractive to consumers \cite{battery_commercialization}. As a result, the market for EVs is expanding rapidly, with more and more automakers introducing electric models and the number of EV charging stations increasing globally \cite{Moradipari_EVCS}. Governments and utility service providers are also promoting the growth of the EV market by providing incentives such as tax credits and grants, as well as investing in the installation of public charging infrastructure \cite{Gov_policy}. The improved performance and increasing availability of EVs contribute to their growing popularity and help accelerate the transition to a low-carbon transportation system \cite{Gov_Incen}.

According to the  Alternative Fuels Data Center (AFDC) report from June 2022 \cite{AFDC}, there are 1,454,480 registered electric vehicles in the US. California has the highest number of EVs, with 563,070, while North Dakota has the lowest, with 380. There is a growing need for public fast charging stations for electric vehicles as the number of EVs on the road continues to increase\cite{Al_EVCS}. Fast charging stations provide a convenient way for EV drivers to recharge their vehicles when they are away from home and are necessary for the widespread adoption of EVs \cite{Ye_EVCS}.  In 2020, it was reported that approximately 12\% of electric vehicle owners did not have access to private charging options, such as a garage or off-street parking with an electrical outlet \cite{Charging_access}. This percentage is estimated to increase to as high as 47\% by 2030, making public charging stations an essential source of power for EV owners. In addition, fast charging stations can help to reduce range anxiety, which is the fear of running out of charge while driving an EV, by providing a quick and convenient way to recharge. The development of a comprehensive network of public fast charging stations can also support the continued growth of the EV market and the transition to a low-carbon transportation system.

The distribution system is a complex network with numerous
varying loads spread in a limited neighborhood \cite{resiliency,VK_Techno}. Connecting distributed generation (DG) to meet the rising and varying demand has been widely researched and implemented \cite{VK_Optimal}. The distributed generation meets the rising demand \cite{VK_P2P}, but the EVs pose a considerable electricity demand for their charging and
operation \cite{Lam_EVCS}. One of the main problems faced is the potential for high levels of power demand, which can strain the local grid and potentially cause power outages. Fast chargers can draw a significant amount of power, often over 100 kW, and the simultaneous charging of multiple EVs can result in a significant increase in power demand. This can lead to voltage dips and fluctuations, which can affect the quality of power supplied to other customers on the distribution system. 
 Optimal placement of electric vehicle charging stations in a distribution system can help to maximize the efficiency and effectiveness of the charging infrastructure \cite{Zeb}. Properly placed charging stations can help to reduce the strain on the distribution system by distributing the demand for charging across multiple locations rather than concentrating it in a single area. This can help to improve the reliability \cite{GNN} and resiliency \cite{DD_VB_Res} of the charging system, as well as reduce the risk of blackouts or other disruptions caused by high demand for charging.

\subsection{Literature Review}
There are several optimization algorithms that can be used to determine the optimal placement of electric vehicle charging stations in an electrical distribution system, including linear programming, mixed integer programming, metaheuristic algorithms, and multi-objective optimization algorithms \cite{EVCS_Optimization}. These algorithms use a variety of techniques to find solutions that maximize or minimize certain objectives, such as cost or coverage, while taking into account constraints such as available locations or demand for charging. Metaheuristic multi-objective optimization algorithms are particularly useful for finding solutions that balance conflicting objectives, such as minimizing cost while maximizing the number of EVs served. These algorithms can help ensure that the charging infrastructure can meet the growing demand for EVs \cite{EVCS_design}.

In \cite{V2G}, fast charging stations are analyzed in a distribution system, but the optimal placement of EVCI is not considered. A mixed integer linear programming (MILP) model for the planning of  plug-in electric vehicle (PEV) fast-charging stations in coupled transportation and  active distribution system (ADS) is proposed in \cite{PEV}, but the stochastic behavior of PEV  arrival/departure and state-of-charge (SoC) are not considered. In \cite{Zeb}, a single objective of minimal power loss is considered using particle swarm optimization (PSO) for optimal sizing and siting of different types of EVCIs in the DS of commercial and residential buildings, including offices and homes. In \cite{Shaaban}, a multi-objective mixed integer non-linear program (MINLP) that jointly specifies optimal locations and sizes of the renewable and non-renewable mix of distributed generation units and EV charging stations is considered for remote islanded microgrids. A multi-objective binary and non-dominated sorting genetic algorithm is used \cite{Asna} to optimally allocate fast single fast chargers at different bus locations in the DS. A single charging station with multiple charging cables at a parking lot with time-of-use (ToU) pricing is analyzed \cite{SOMC}, but the location and impact of this EVCI at the DS are not considered. The multi-port EVCSs for fixed charging rates are analyzed, and even zero-price for charging is promoted \cite{Free}. The design and planning of a multi-charger with a multi-port charging system for PEV charging station are discussed \cite{Chen_Multi_port}, but the impact on distribution is not analyzed. 
A hybrid of grey wolf optimization and particle swarm optimization is used \cite{Bilal} for optimal allocation of 2 EVCIs with 1500kW rating each for the IEEE 33-bus distribution systems, but pricing mechanisms are not analyzed. A hybrid method based on the MOPSO optimization algorithm and sequential Monte Carlo simulation is presented \cite{Hadian}, and the EVs arrival and departure times are considered using a probability distribution function. In \cite{Tounsi}, a hybrid bacterial foraging optimization algorithm and particle swarm optimization (BFOA-PSO) technique is proposed for the optimal placement of EVCIs into the distribution network; 11 kW and 22 kW chargers are considered for charging five different types of EVs. A hybrid chicken swarm optimization and teaching learning-based optimization algorithm is utilized to obtain the Pareto optimal solution \cite{Deb} for placement of 3 EVCIs; fast charging is not considered. A hybrid optimization algorithm (PSO-DS) combining particle swarm optimization algorithm and direct search method is used \cite{Muthukannan}, but fast chargers are not considered. The optimal siting and sizing problem of the fast charging station is solved by the Voronoi diagram together with the particle swarm optimization algorithm \cite{Zhang_H}, but the impact on the distribution system is not analyzed. In most of the above works, time-varying analysis is not performed considering random EV state of charge and arrival/departure time. The incorporation of dynamic price models and cost predictions are not analyzed.

Dynamic pricing for electric vehicles refers to a pricing strategy in which the cost of charging an EV is varied based on demand and supply conditions \cite{Price_1}. This approach aims to incentivize EV owners to charge their vehicles when electricity demand is low, thereby helping to balance the grid and reduce the need for expensive peak power generation \cite{Price_2}. Dynamic pricing for electric vehicles can be implemented through time-of-use pricing, where the cost of charging varies based on the time of day, or through real-time pricing, where the cost is adjusted based on the current demand and supply conditions \cite{Price_3}. By using dynamic pricing, utilities can better manage their electricity supply and demand, and EV owners can potentially save money on their charging costs \cite{Price_4}. In all of these works, the benefits of dynamic price are presented, but the benefits of predicting the price are not discussed.

\subsection{Main Contributions and Paper Organisation}

The central contribution of the paper is the optimal placement of fast electric vehicle charging infrastructure achieved using a multi-objective particle swarm optimization algorithm. This involves finding the best possible locations for the placement of multiple EVCI such that the distribution system has minimal power loss and voltage deviation. A time-varying analysis of the EVCI in the distribution is performed. We also predict the dynamic price for electric vehicle charging that improves efficiency, increases revenue, enhances the user experience, and improves grid stability. By predicting prices in real-time, charging station owners and operators can better manage their energy consumption, optimize the use of renewable energy sources, and adjust prices based on predicted demand and other factors. Additionally, predicting prices in real-time and adjusting charging patterns accordingly can help to stabilize the grid and reduce the risk of blackouts or other disruptions caused by high demand for charging. Overall, predicting the dynamic price for EV charging using data analytics \cite{MP_USpatent,MP_WOpatent} and machine learning methods \cite{MP_arxiv,DD_arxiv} can help to optimize the use of charging infrastructure and support the continued growth of the EV market.

Salient features of the proposed work that differentiate it from similar approaches described in existing literature are as follows:

\begin{enumerate}
    \item A strategy for allocating five fast-charging EVCIs in a distribution system using a modified multi-objective particle swarm optimization algorithm, with the goal of minimizing power loss and voltage deviation in a distribution system, ensuring a globally optimal solution.
    \item A model that takes into account time-varying loads and EV charging patterns, including random arrival times and states of charge, as well as a real-time pricing system based on utility prices and EV traffic to maximize profits.
    \item An EV price prediction model using the ARIMA algorithm allows EVCI operators to make promotional offers at different periods to increase profits and utilization of the EVCIs.
\end{enumerate}

The rest of the paper is organized as follows; Section \ref{section:formulation} presents the EVCI allocation problem and the methodologies used for the optimal allocation of EVCI. Section \ref{section:MOPSO} presents in detail the steps followed in the MOPSO algorithm. In Section \ref{Section:Prediction}, the price model of EVCI and the cost prediction method are discussed. In Section \ref{section:Results}, we discuss simulation models and analyze the results for time-varying data. Finally, conclusions are drawn in Section \ref{section:Conclusion}.

\section{Problem Formulation and Methodology}
\label{section:formulation}
\subsection{EVCI Allocation Problem}

\begin{table}[b]
\centering
\caption{EVCI Parameters.}
\begin{tabular}{lc}
\toprule
\multicolumn{1}{c}{Parameters} & Values   \\
\midrule
No. of EVCIs                    & 5        \\
No. of EVCSs / EVCI            & 20       \\
Total No. of EVCSs             & 100      \\
No. of EVs / EVCS / Day        & 6-10     \\
No. of EVs / EVCI/ Day         & 120-200  \\
Total No. of EVs / Day         & 600-1000 \\
Charger Rating (kWh)           & 50      \\
\bottomrule
\end{tabular}
\label{tab:EV_par}
\end{table}

\begin{figure}[t]
  \centering
  \includegraphics[width=3.25in]{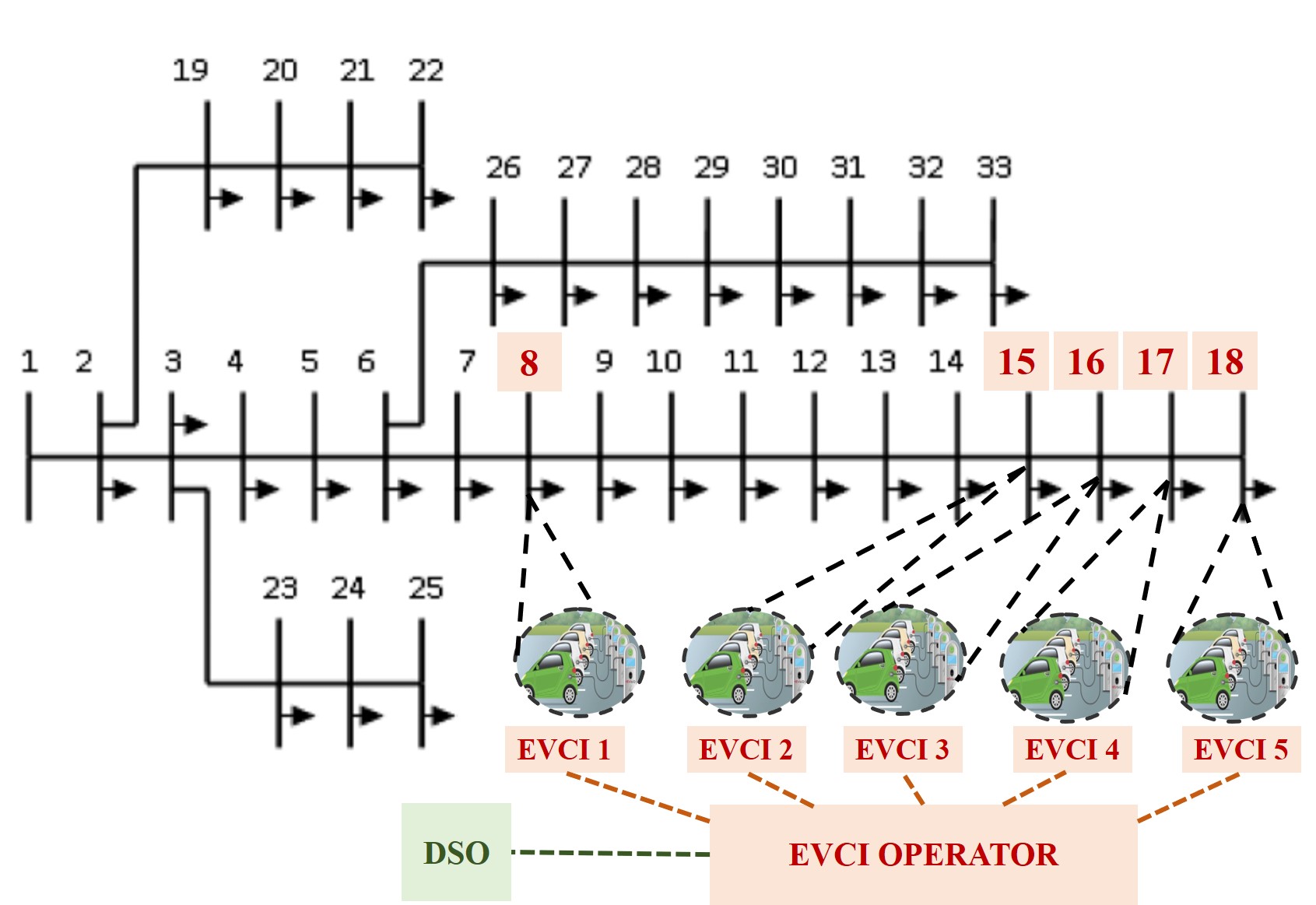}
  \caption{IEEE 33 bus distribution system with electric vehicle charging infrastructures at 5 different bus locations.}
  \label{fig:schematic}
\end{figure}

The allocation problem involves two main factors: the position of the EVCI within the DS and the load capacity of each EVCI.  We consider to allocate 5 EVCIs at different bus locations in the distribution system. A schematic of a DS integrated with EVCIs is shown in Fig. \ref{fig:schematic}. We consider each EVCI to have 20 charging stations, each with a rating of 50 kW, for a total load capacity of 1000 kW. This is based on the assumption of a growing EV fleet, as well as the need for fast charging options to accommodate increasing demand, support long-distance travel, and cater to users without access to private charging infrastructure. Table \ref{tab:EV_par} lists the parameters of each EVCI and the number of EVs expected to arrive at the EVCIs. Table \ref{tab:EV_cap} shows the details of the five different types of EVs, with varying battery capacities, that we have considered arriving at the EVCIs in a day. We can allocate the EVCI as a load on the DS at different locations to benefit both the DS and the EVCI. We need to analyze the impact of the EVCI on the DS at various locations (assuming that all the buses except the substation are feasible locations for the EVCI) and make an optimal decision in terms of voltage deviation and power loss.

We can define the problem of computing the location of EVCIs as follows:
\begin{itemize}
    \item Objectives: Minimizing power loss and voltage deviation
    \item Fixed variables: Load capacity of EVCIs, load on DS.
    \item Decision variables: Location of 5 EVCIs 
    \item  Method/algorithm: Iterative search method and multi-objective particle swarm optimization.
\end{itemize}

\begin{table}[H]
\centering
\caption{Battery Capacity of EV Models Considered.}
\begin{tabular}{lc}
\toprule
\multicolumn{1}{c}{EV Model}       & Battery Capacity (kWh) \\
\midrule
Nissan leaf                        & 24                     \\
Nissan e-NV200                     & 40                     \\
Tesla model 3 standard plus saloon & 55                     \\
Tesla model 3 long range saloon    & 75                     \\
BYD e6                             & 82           \\
\bottomrule
\end{tabular}
\label{tab:EV_cap}
\end{table}

\subsection{EVCI Placement Methods}

\subsubsection{Iterative Search Method} 

This method involves a time-consuming approach of meticulously computing the power loss and voltage deviation for each possible combination in the search space. The total number of combinations for placing ${N_{evci}}$ number of EVCI at ${(N_{bus}-1)}$ buses of the distribution system will be $^{(N_{bus}-1)}C_{(N_{evci})}$, as EVCI cannot be placed at the first bus which is the substation of the DS.

\vspace{6pt}
\hrule
\vspace{3pt}
\textbf{Pseudocode for Iterative Search Method:}
\vspace{3pt}
\hrule
\vspace{6pt}

BEGIN

INITIALIZE number, size and possible location of EVCIs

GENERATE list of all possible location combinations 

FOR each combination

	\hspace{5mm}INPUT line and load data of DS
	
	\hspace{5mm}ADD EVCIs as load 
	
	\hspace{5mm}UPDATE load power at the EVCI locations
	
	\hspace{5mm}PERFORM load flow 
	
	\hspace{5mm}COMPUTE objective functions

END FOR
	
PLOT objective function values for all combinations

COMPUTE solution with minimum objective function

DISPLAY best solution

END

\vspace{5pt}
\hrule
\vspace{6pt}

Although this method of computing the best location of 5 EVCIs provides a global optimal solution by considering all possible solutions in the entire search space, it takes a huge amount of time and can be even more burdensome for larger systems with more EVCIs. To address the issue of time and computational burden, we can use stochastic or heuristic-based optimization algorithms.

\subsubsection{Multi-Objective Particle Swarm Optimization} 

Particle swarm optimization is a metaheuristic optimization algorithm that is inspired by the behavior of a swarm of particles, such as birds or insects and is often used to find the global optimum of a function in a search space. Single objective PSO algorithms arrive at a single global best solution. However, multi-objective optimization using PSO can be performed by either using the weighted sum optimization method or by achieving multi-objective Pareto optimality \cite{TJ}. Weighted-sum multi-objective particle swarm optimization is a variant of the multi-objective particle swarm optimization algorithm that converts a multi-objective optimization problem into a single-objective optimization problem. In this method, weights are assigned to the objective functions, and the optimization process is performed by minimizing the weighted sum of the objectives. This approach is useful when the decision maker clearly prefers one of the objective functions and wants to give it more importance in the optimization process. However, it may not always result in a global optimal solution, as the solution is largely determined by the weights assigned to the objectives. MOPSO based on Pareto optimality is designed to find a set of non-dominated solutions (also known as the Pareto front) for a multi-objective optimization problem, rather than a single global optimum. It considers multiple conflicting objectives simultaneously and generates a set of non-dominated solutions that trade-off between them. We can identify the best-compromised solution as the optimal solution from this set of solutions. Since the search for optimality is not constrained in this method, we use it for our EVCI allocation problem. 

\begin{figure*}[h!]
  \centering
  \includegraphics[width=6.5in]{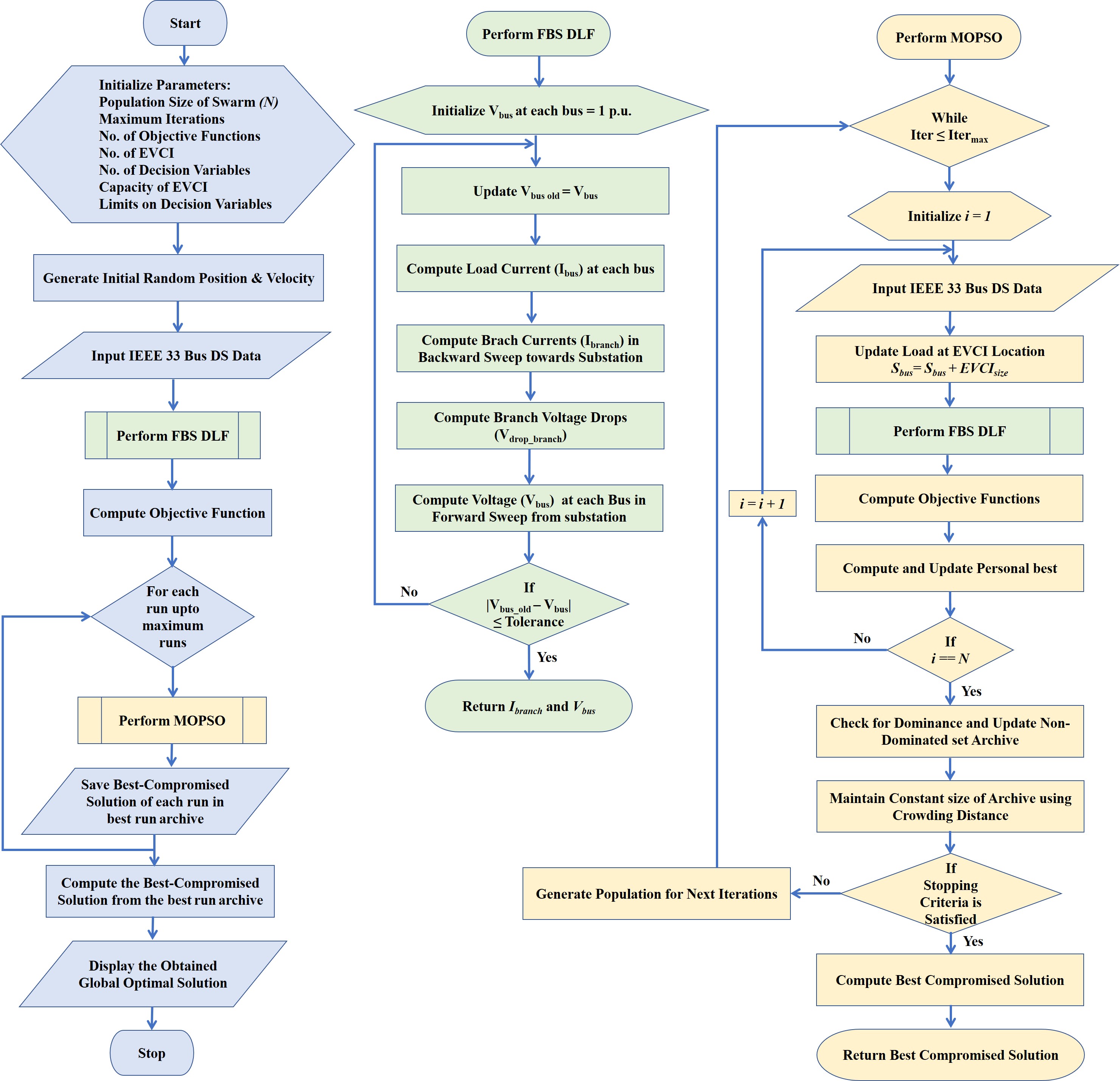}
  \caption{Flowchart of allocation of EVCI using MOPSO.}
  \label{fig:flowchart}
\end{figure*}

\section{EVCI Allocation using MOPSO}
\label{section:MOPSO}

\subsection{Proposed Modified MOPSO Algorithm}

The detailed flowchart of the allocation of EVCI in the IEEE 33-bus system is shown in Fig. \ref{fig:flowchart}, the main program with two functions is shown; i) Forward-backward sweep (FBS) used to perform distribution load flow (DLF) analysis and ii) proposed modified MOPSO. This section elaborates on the key steps in the proposed method.

\subsubsection{Decision variables}

The problem involves finding the locations for 5 EVCIs in a DS. The size of each EVCI is fixed at 1000 kW, and they can be placed at any of the buses in the system except the substation. The decision variable for this problem is the location of each EVCI, which can take on integer values from 2 to $N_{bus}$.

\vspace{6pt}
\subsubsection{Optimization objectives}

While searching for the optimal locations for the EVCIs in the DS, one of our aim is to minimize the power loss in the branches. The total active power loss in the DS can be calculated as follows,

\begin{equation}
\label{eqn:min_P}
    \textup{min}\sum_{branch=1}^{n \;branch} 
    I^2 _{branch} . R_{branch}
\end{equation}

where, $I_{branch}$ and $R_{branch}$ are the current and resistance in the branch, respectively. Minimizing this quantity is the first objective of the optimization problem.

In addition to minimizing power loss in the DS, it is also important to maintain a constant voltage level at all nodes/buses in the system. The permissible voltage deviation in the DS is $\pm 5\%$. The voltage deviation at a bus can be calculated using the equation:

\begin{equation}
V_{bus_{dev}} = V_{threshold} - V_{bus}
\end{equation}

where $V_{threshold}$ is the desired threshold voltage and $V_{bus}$ is the voltage at the bus. Minimizing the squared voltage deviation at the buses improves the voltage profile and keeps it within acceptable limits. The second objective of the optimization problem is given by:

\begin{equation}
\label{eqn:min_V}
     \textup{min}\sum_{bus=2}^{n \;bus} V^2 _{bus_{dev}}
\end{equation}

\subsubsection{Personal best archive}

For each particle, we calculate and store the best location (variable) in terms of the minimum objective functions (loss and voltage deviation) in a personal best archive. If a new personal best value is found, we update the archive with the particle's new position.

\subsubsection{Non-dominated set archive (Pareto optimal front)}

We need to find solutions that optimize both minimum power loss and minimum voltage squared deviation. However, this can lead to contradictory results. To address this issue, we aim to find non-dominated solutions, which are solutions that are not dominated by any other solution in terms of both objective functions. To update the non-dominated solutions in the archive, we perform the following steps:
\begin{enumerate}[i)]
  \item Check if the fitness value of archive member $a$ is dominated by solution $s$.
  \item If $a$ is dominated by $s$, it is replaced by solution $s$.
   \item If $s$ is dominated by $a$, move on to the next solution by setting $s=s+1$.
   \item If there is no domination between $s$ and $a$, both $s$ and $a$ are kept in the archive.
  \end{enumerate}

\subsubsection{Archive size}

As the number of iterations increases, the number of non-dominated solutions in archive $A$ also increases, leading to computational difficulties and longer processing times. To address this issue, we limit the size of the archive by using crowding distance. This is done as follows:

\begin{enumerate}[i)]
  \item Initialize the distance of each particle,  $d^k=0$
  \item Set the distance $d^k=\infty$ for boundary solutions, and calculate the distance $d^k$ for the remaining solutions using the equation:

\begin{equation}
  d^k=d^k+\frac{f^{k+1}_{m} - f^{k-1}_{m}}{f^{max}_{m} - f^{min}_{m}}  
\end{equation}

 where $f$ is the objective function value, $k$ is the considered particle, $m$ is the considered objective, $f_m^k$ is the value of $m^{th}$ objective of the $k^{th}$ particle of the archive $A$, $f_m^{min}$ is the minimum objective value, and $f_m^{max}$ is the maximum objective value.
 
  \item Sort the distances in descending order and select the first $p$ solutions for the next iteration.
\end{enumerate}

This way, the size of the archive is limited to a maximum of $A_{max}$ non-dominated solutions.

\subsubsection{Position and velocity update}

In single-objective optimization, a single global best position is used to guide the entire swarm in updating its next position and velocity. However, using this approach for multi-objective optimization might lead to the loss of diversity of solutions from the Pareto front. To address this, we compute a local best guide, denoted as $L_{gb}$, for each particle, which serves as its global best. The local best guide is determined using the Sigma method as follows:
\begin{enumerate}[i)]
  \item Calculate the sigma values, $\sigma_k$, for each particle, and $\sigma_l$ for each archive member.
  \item Calculate the distance $dist=\sigma_k -\sigma_l$
  \item The archive member that gives the least distance for particle $k$ is its local best guide.
\end{enumerate}
 The position $x_k$ and velocity $v_k$ of each particle $k$ are then updated  based on its personal best and local best guide, as shown in the following equations:
\begin{multline}
    v_k(t) = v_k(t-1)+c_1.r_1.(P_{b_k}-x_k(t-1))\\
    +c_2.r_2.(L_{gb_{k}}-x_k(t-1))
\end{multline}
\begin{equation}
    x_k(t)=x_k(t-1)+v_k(t)
\end{equation}
 where $c_1$ and $c_2$ are inertia constants, $P_{b_k}$ is the personal best of particle $k$, $L_{gb_k}$ is the best local guide of particle $k$, $r_1$ and $r_2$ are random values that control the movement of the particle towards the best optimal solution.

\subsubsection{Best-compromised solution}

We have several non-dominated solutions on the Pareto front. To select the best-compromised solution, we use fuzzy membership functions as follows:

\begin{equation}
\mu_m^{k}=\left\{
    \begin{aligned}
        & 1 \qquad  \qquad \qquad \textup{if} \quad f_m \leq f_m^{\textup{min}} \\
        & \frac{f_m^{\textup{max}}-f_m}{f_m^{\textup{max}}-f_m^{\textup{min}}} \quad \; \, \textup{if} \quad f_m^{\textup{min}} < f_m < f_m^{\textup{max}} \\
        & 0 \qquad \qquad \qquad \textup{if} \quad f_m \geq f_m^{\textup{max}} \\
    \end{aligned}\right.
\end{equation}

where $\mu_m^{k}$ is the fuzzy membership value of the objective $m$ for particle $k$. The membership function of the solution for each particle $k$ of the archive $A$ is normalized and can be represented as follows:

\begin{equation}
    \mu^k_{norm}=\dfrac{\sum\limits_{m=1}^{N_{obj}}  \mu_m^k}{\sum\limits_{k=1}^{A} \sum\limits_{m=1}^{N_{obj}} \mu_m^k}
\end{equation}

where $N_{obj}$ is the number of objectives. The Pareto optimal solution that has the highest $\mu^k_{norm}$ value is considered as the best-compromised solution.

\subsubsection{Stopping criterion}

The stopping criterion for the EVCI allocation using MOPSO is based on the number of times the same solution is repeated. If the same solution is identified as the best compromise for $i$ consecutive runs, the MOPSO is considered to have converged. However, each execution of the code resulted in different optimal solutions, indicating that the algorithm has not reached its global optimal solution, but rather has settled for a local optimal solution.

To address this issue and ensure global optimality, the stopping criterion was modified to include multiple runs of the same algorithm. The entire MOPSO code is repeated for a maximum number of runs. After the convergence of each run, the best solution is stored in a separate best-run archive $A_R$. The best-compromised solution from the best-run archive is then computed and displayed as the global optimal solution.

The global optimality of the final solution is justified by the following:

\begin{enumerate}[-]
    \item The same optimal solution is obtained after several executions of the entire code with different random initial populations
    \item The solution obtained using the modified MOPSO is the same as the one obtained from the iterative search method.
\end{enumerate}

\section{Dynamic Price Model and Prediction}
\label{Section:Prediction}

\subsection{EVCI Price Model}
We consider that the grid has a dynamic pricing system, where the price of buying energy from the grid varies on an hourly basis. The distribution system operator varies this price and informs the EVCI operator of the price variations every hour. Based on this price variation, the EVCI operator needs to accordingly vary the price of selling energy to the EV customers at the charging station. The EVCI charging fee is based on two prices; one is a fixed price $R_f$ and the second is a time-based price $R_t$ for each hour in time $t$, the total price set by the EVCI operator is given by
\begin{equation}
    EVCI_{price} = R_{f} + R_{t}
\end{equation}
% \begin{equation} \label{eq:R_p}
% R_=\left\{ 
%     \begin{aligned}
%         & R_p \quad \; \textup{if} \quad t=p\\
%         & R_n \quad \; \textup{if} \quad t=n\\
%         & R_{op} \quad \textup{if} \quad t=op
%     \end{aligned}\right.
% \end{equation}
The fixed price $R_f$ is a constant price for all EVs at any time of day; this is a standard service fee. The time-based price $R_t$ depends on the EV traffic conditions and daily varies for three periods: $R_p$ for peak $p$ hours, $R_n$ for normal $n$ hours, and $R_{op}$ for off-peak $op$ hours. The daily cost of the EVCI is given by
\begin{multline}\label{eq:costwithSandB}
    EVCI_{cost\slash day} = \sum_{i}^{N_T} R_f EV_i + \sum_{i}^{N_p} R_p EV_i + \sum_{i}^{N_n} R_n EV_i \\ + \sum_{i}^{N_{op}} R_p EV_i
\end{multline}
where the total number of EVs at all charging stations at an EVCI for a day is $N_T=N_p+N_n+N_{op}$ and $N_p$, $N_n$ and $N_{op}$ are the total number of EVs in a day during peak, normal and off-peak hours respectively.

\subsection{EVCI Price Prediction}

Once we have the EVCI price model, we can collect the EVCI price data for a period of time. Based on previous data we can make predictions of the price so that the EVCI operator can make promotional offers and increase utilization and cost benefits. For the prediction of EVCI price, we use the autoregressive integrated moving average (ARIMA) method for predicting time-series data. It incorporates autocorrelation measures within the time series data to forecast future values. The model's autoregression part measures a particular sample's dependency with a few past observations. These differences are measured and integrated to make the data patterns stationary or minimize the obvious correlation with past data. The general form of the ARIMA $(p,d,q)$ model \cite{Book_stat} is,
\begin{equation}
\Phi_p(B) \nabla^d X_t=\Theta_q(B)\varepsilon_t
\end{equation}

where $\varepsilon_t$ is a random shock with mean zero and var ($\varepsilon_t$) = $\sigma^2_\varepsilon$. We indicate
lagged observations with the backshift operator $B$, defined to mean $BX_t=X_{t-1}$. The conventional notation for a time series is $X_t=1,2,...,n$. $\Phi_p(B)=\phi(B)=1-\phi_1B-...-\phi_pB^p$ is a $p^{th}$ degree polynomial and $\phi_p$ is the correlation with lag $p$. $\Theta_q(B)=\theta(B)=1-\theta_1B-...-\theta_q B^q$ is a $q^{th}$ degree polynomial and $\theta_q$ is the correlation with lag $q$. The differencing with order $d$ is represented by $\nabla^d$.

Determining the parameters $p$, $d$, and $q$ for implementing ARIMA models is crucial. We can use the augmented-dickey fuller test (ADF), autocorrelation function (ACF), and Partial autocorrelation function (PACF) to determine our model parameters. The value of differencing $d$ is found by using ADF. Based on the values of PACF for different orders of differencing, we can obtain the value of $p$; observing the values of ACF for different orders of differencing, we can obtain the value of $q$. 

\section{Simulation Results and Discussion}
\label{section:Results}

In this work, we examine the optimal placement of five EVCIs in a 33-bus radial DS with a voltage of 12.66 kV and load size of 3.715 MW. The system consists of 33 buses and 32 lines, as shown in Fig. \ref{fig:schematic}. The EVCIs, each equipped with 20 fast charging stations that can charge an electric vehicle (EV) at a rate of 50 kW, are placed at five different locations in the system.

\subsection{MOPSO Results}

\begin{figure}
  \centering
  \includegraphics[width=3.25in]{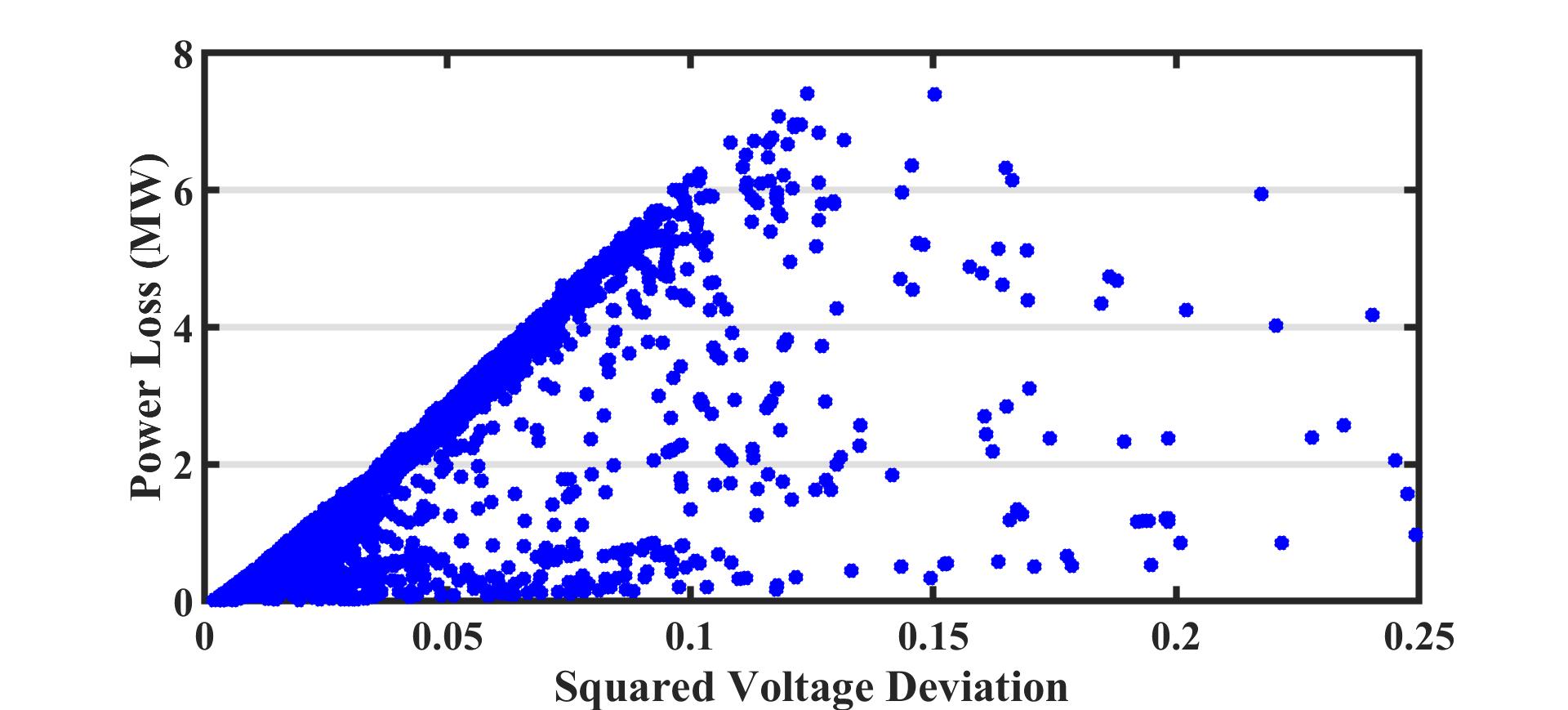}
  \caption{Entire search space of the EVCI allocation problem obtained from the iterative search method with each point corresponding to a set of 5 different locations of EVCIs.}
  \label{fig:search_space}
\end{figure}

The MOPSO model was created using MATLAB to simulate a multi-objective optimization problem involving the placement of five EVCIs in a power DS. The maximum rated load for each EVCI is set at 1000 kW, and the goal is to minimize power loss and voltage deviation across the entire system by finding the optimal locations for the EVCIs. To find these locations, we first use the iterative search method where the EVCIs are placed at all possible locations and a load flow analysis is performed to calculate the power loss and voltage deviation for each configuration. In the case of an IEEE 33-bus system, there are 201,376 possible combinations to consider, as shown in Fig. \ref{fig:search_space}. The optimal locations can then be selected based on the least power loss and voltage deviation values. However, this method can be computationally intensive, especially for larger systems with more constraints.

\begin{figure}
  \centering
  \includegraphics[width=3.25in]{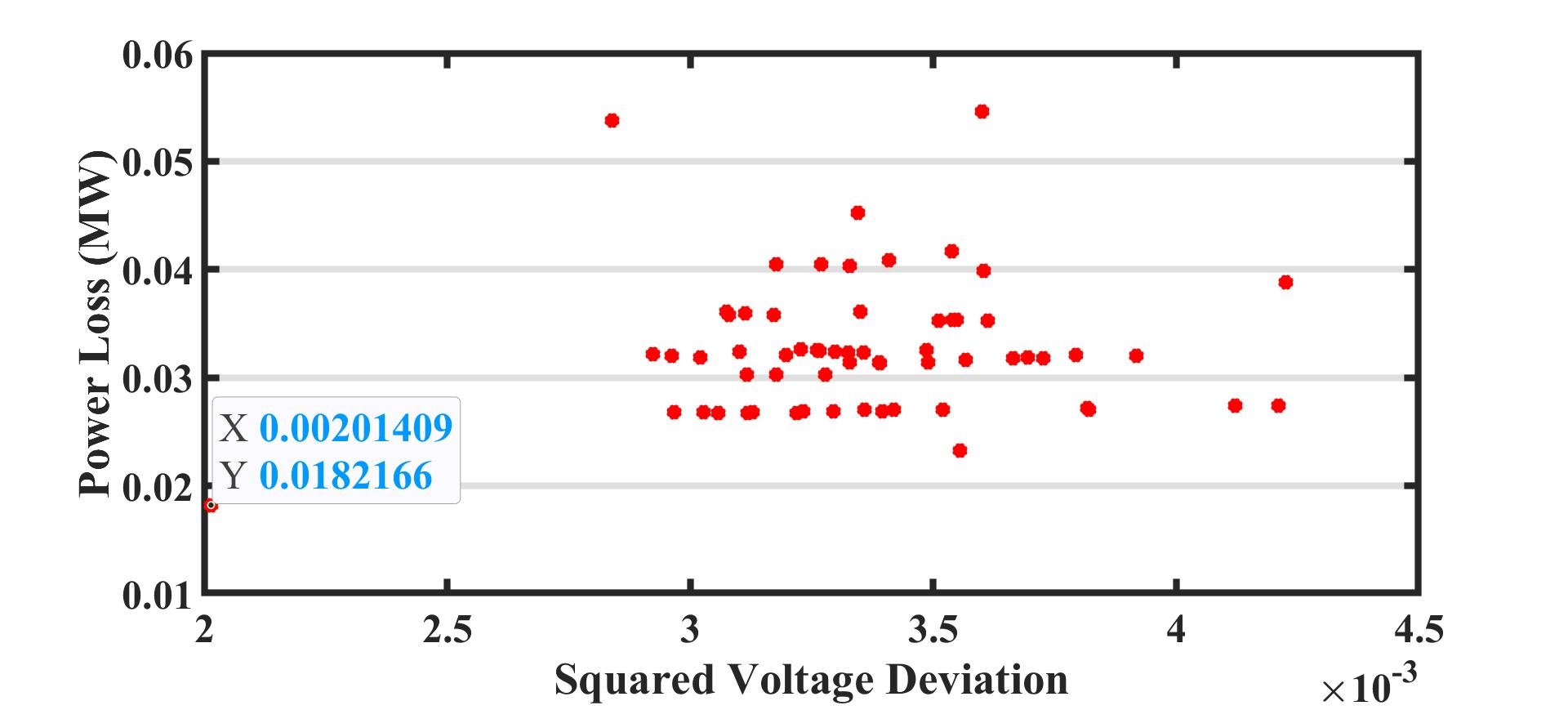}
  \caption{Best compromised solutions obtained from each run of the MOPSO algorithm with each point corresponding to the solution of a set of 5 different locations of EVCIs.}
  \label{fig:best_solutions}
\end{figure}

\begin{figure}
  \centering
  \includegraphics[width=3.25in]{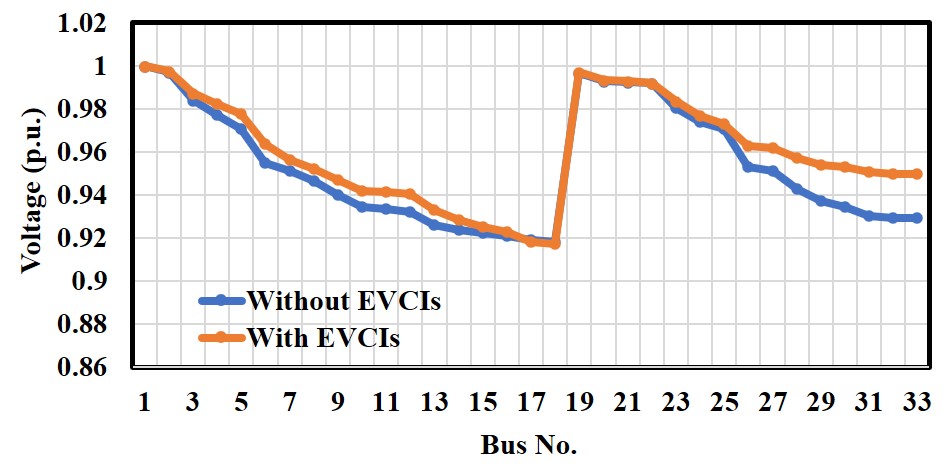}
  \caption{Voltages at each bus in the distribution system for two cases i) without EVCIs ii) with EVCIs.}
  \label{fig:mopso_voltages}
\end{figure}

\begin{figure}
  \centering
  \includegraphics[width=3.25in]{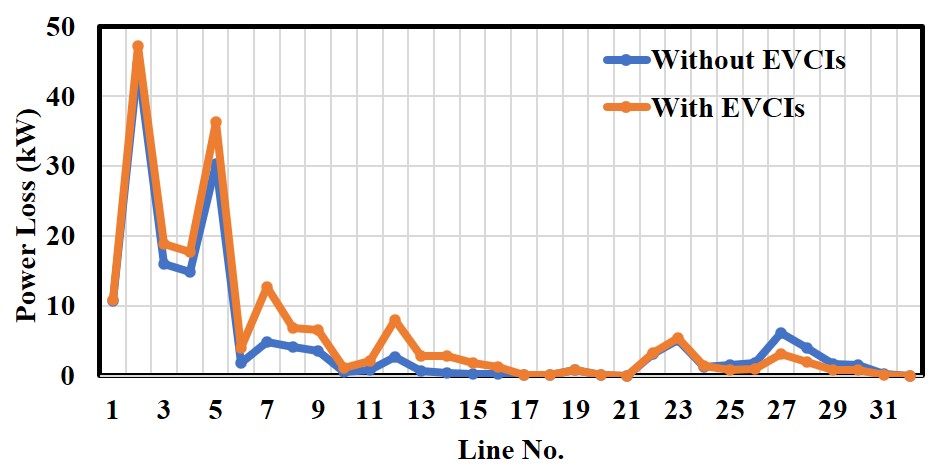}
  \caption{Power loss in each line of the distribution system for two cases i) without EVCIs ii) with EVCIs.}
  \label{fig:mopso_losses}
\end{figure}

The proposed solution in this study is the use of a multi-objective particle swarm optimization algorithm to determine the optimal placement of EVCIs while taking into account power loss and squared voltage deviation objectives. The computation time for one single run with a population of 500 for the MOPSO method and the proposed modified MOPSO method is 37.59 seconds and 20.24 seconds, respectively. For both methods, convergence was achieved at 51 iterations. The results of the MOPSO algorithm, presented in Section \ref{section:MOPSO}, are illustrated in Fig. \ref{fig:best_solutions}. From the best-compromised solutions obtained from each run, it can be observed that the solution closest to zero has a minimum power loss of 201.40 kW and a minimum squared voltage deviation of 0.0182 p.u. This solution is the final best compromised global optimal solution obtained from the proposed method. It is also same as the best solution obtained from the iterative search method; thereby, validating the proposed method. The EVCI locations corresponding to this solution are at buses 8, 15, 16, 17, and 18. The lowest voltage found was at bus 18 with a value of 0.9172, as shown in Fig. \ref{fig:mopso_voltages}, and the lowest power loss was on Line 32 with a value of 0.0065 kW, as shown in Fig. \ref{fig:mopso_losses}.  The optimal placement of EVCIs based on the considered two objectives would help in achieving a more balanced load distribution, thus resulting in improved voltages at certain buses. The results of the MOPSO algorithm are compared to the base case in Table \ref{tab:mopso_results}.

It is to be noted that we have not considered the spread of EVCIs in the DS to meet the practical considerations of the transportation network as considered in \cite{PEV}. If the transportation network is considered, the locations of the EVCIs would be preferred at different buses which would lead to higher power loss and increased voltage deviations. Thus a trade-off needs to be made, by considering the maximum allowable limit for power loss and voltage deviations of the DS.

\begin{table}
\centering
\caption{Comparative results for with and without EVCIs}
\begin{tabular}{lcc}
\toprule
Parameter & Without EVCIs & With EVCIs \\
\midrule
Optimal Locations of 5 EVCIs & - & 8,15,16,17,18 \\
Power Loss (kW) & 164.36	&201.40 \\
Squared Voltage Deviation (p.u.) & 0.0235	&0.0182  \\
Min Voltage at Bus (p.u.) & 0.9183 (Bus18)	&0.91729 (Bus18) \\
Min Power loss in line (kW) &0.0128 (Line 32)	& 0.0065 (Line 32) \\
\bottomrule
\end{tabular}
\label{tab:mopso_results}
\end{table}

\subsection{Time-Series Analysis}

\begin{figure}
  \centering
  \includegraphics[width=3.25in]{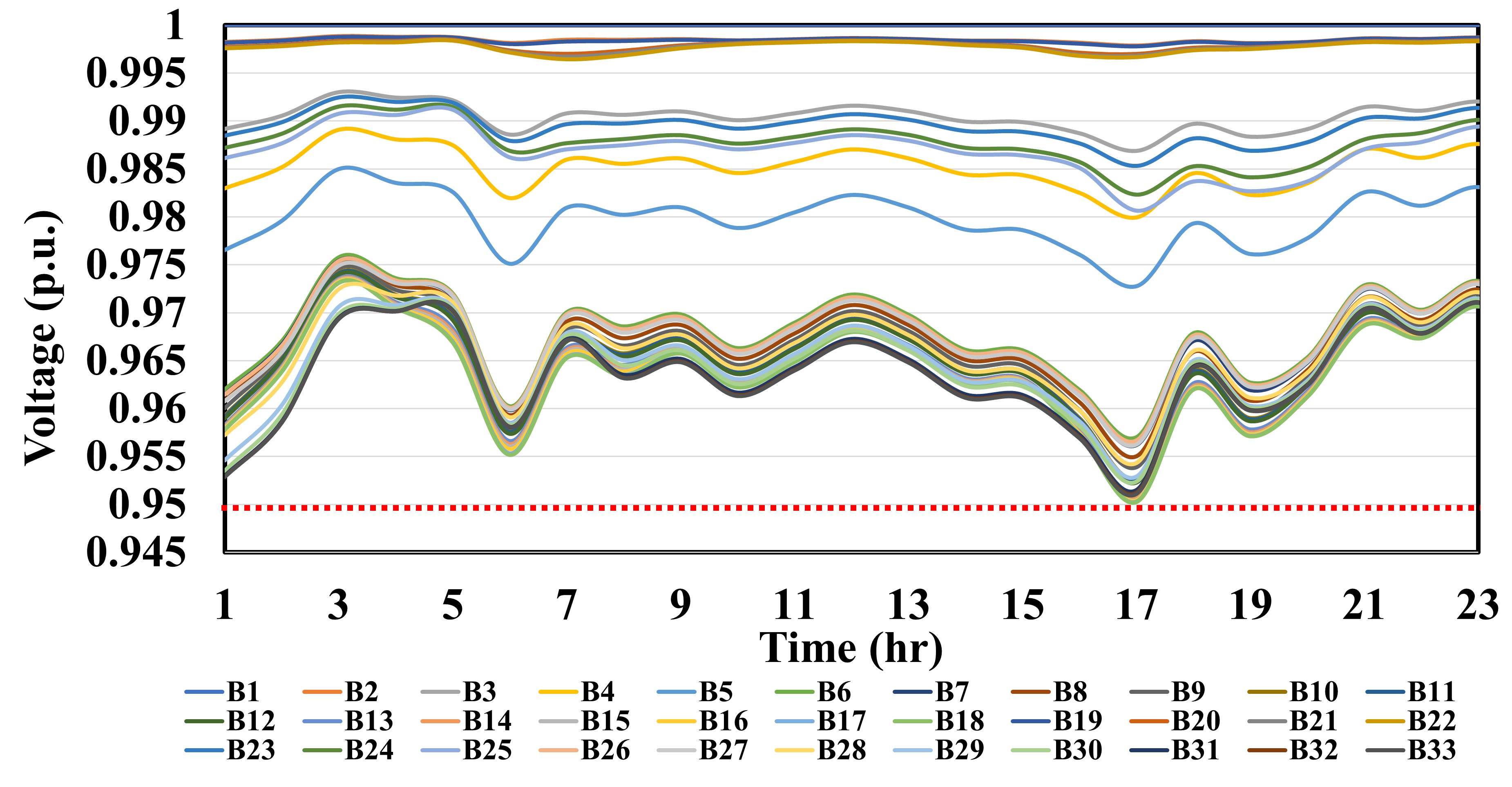}
  \caption{Voltages at each bus in the distribution system for Day 1 with EVCIs installed at 5 buses}
  \label{fig:bus_voltage}
\end{figure}

\begin{figure}
  \centering
  \includegraphics[width=3.25in]{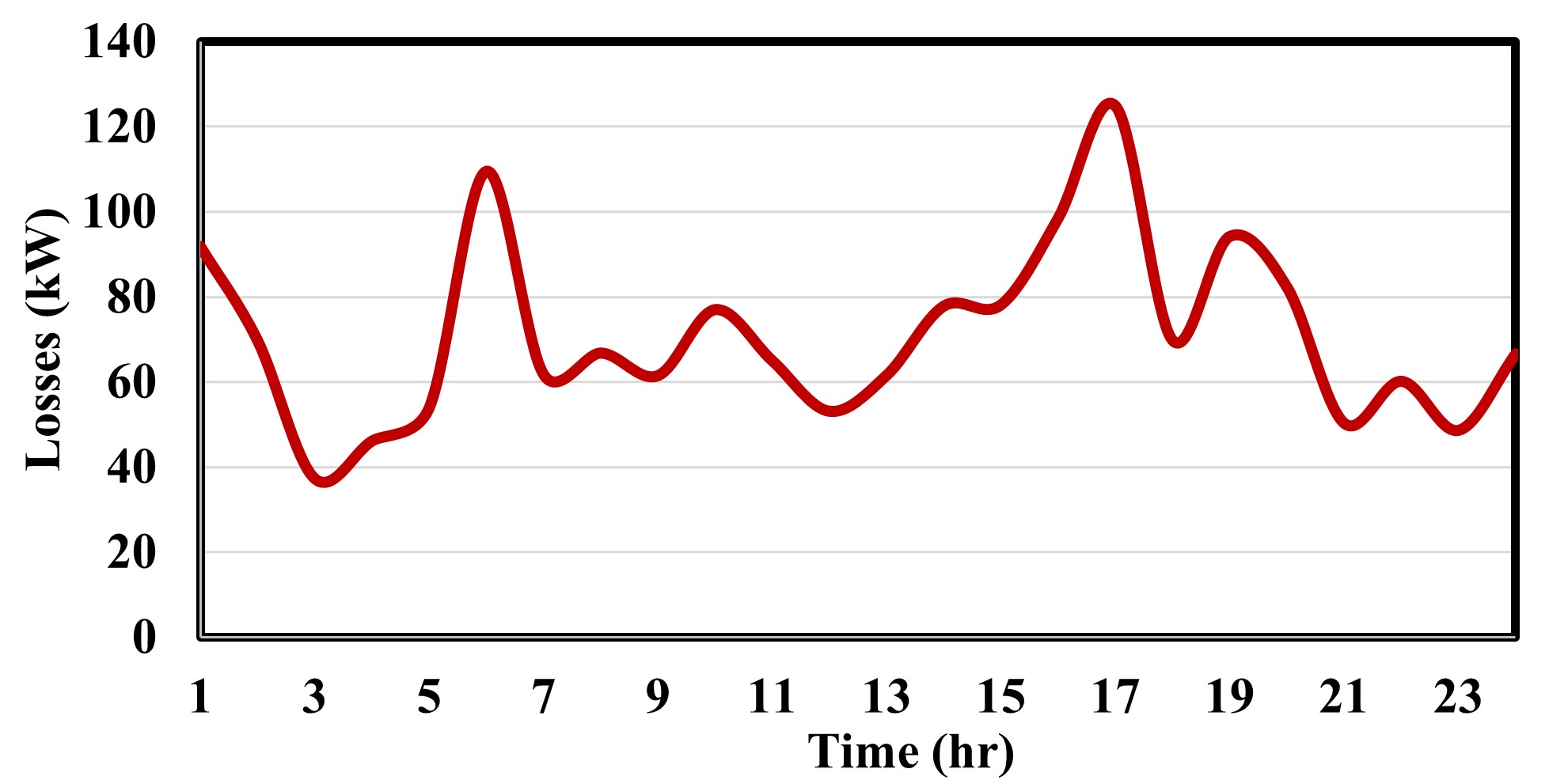}
  \caption{The total system power losses in the distribution system for Day 1 with EVCIs installed at 5 buses}
  \label{fig:system_losses}
\end{figure}

\begin{figure}
  \centering
  \includegraphics[width=3.25in]{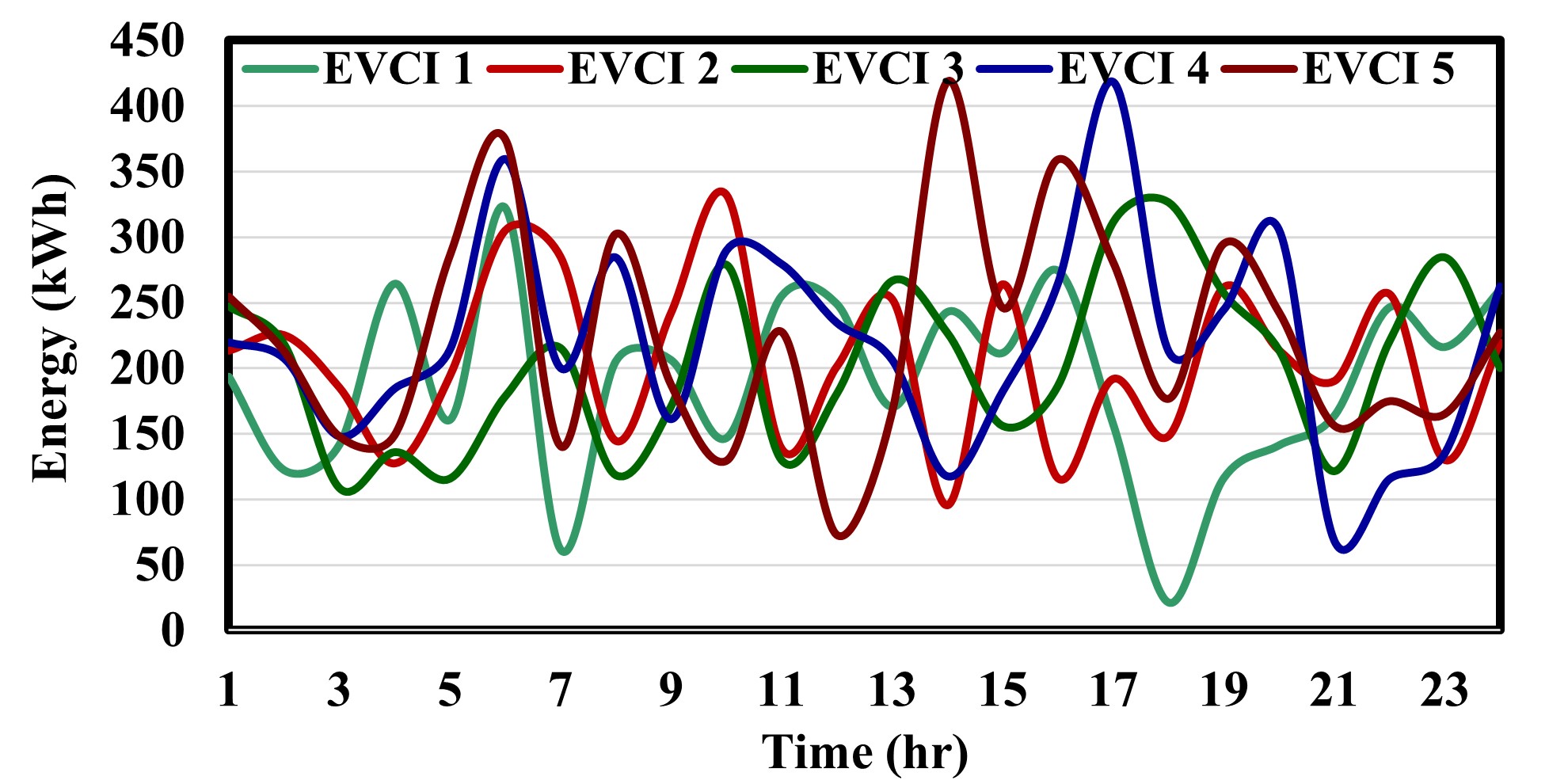}
  \caption{Energy of all 5 EVCIs connected to the distribution system with varying EV charging load for Day 1.}
  \label{fig:EVCS_energy}
\end{figure}

Now that we have computed the locations of the five EVCIs at different buses in the distributions system. 
We perform a time series analysis considering time-varying loads at the buses. For this purpose, an hourly load profile dataset from DTU data \cite{dataset} is used. We consider the five EVs listed in Table II as time-varying EV loads at each charging station. These EVs could be charged in under one hour by using the 50 kW fast charger. Simulation of data for EVs at charging stations considering various real-time societal conditions like road traffic, weather, public holidays, weekends, weekdays, night hours, peak hours, off-peak hours, electricity prices, and EV adoption is practically difficult, and such real-time EV data for research purposes is barely available. It is worth emphasizing that the proposed pricing model and the price prediction method could accommodate such variations of real-world EV data scenarios and demonstrate similar results and accuracy. In this work, time-series analysis is performed using the simulated data generated under the assumption that 6-10 EVs would arrive at each charging station with random arrival times within a 24-hour period which helps to approximate the data with real-world conditions. The state of charge for each EV's battery bank arriving at the charging station is also randomly considered to be between 20\% and 80\%. This assumption is based on a real-world study of the optimal charging for electric vehicles \cite{SoC_limits}. Once charged to the maximum state of charge (80\%), the EVs are expected to depart from the EVCIs. We have generated hourly data for 8 months, i.e., January to August, considering the available load profile dataset. We now perform the time-series simulation using OpenDSS supported by MATLAB. We show the analysis for the first 24 hrs. in Fig. \ref{fig:bus_voltage}; we can observe that the voltages are within limits of 0.95 p.u. and the total power loss in the system is low in Fig. \ref{fig:system_losses}. The total energy of all the 5 EVCIs with charging stations considered together is presented in Fig. \ref{fig:EVCS_energy}; they are random for a 24hr duration. 

\begin{figure}
  \centering
  \includegraphics[width=3.25in]{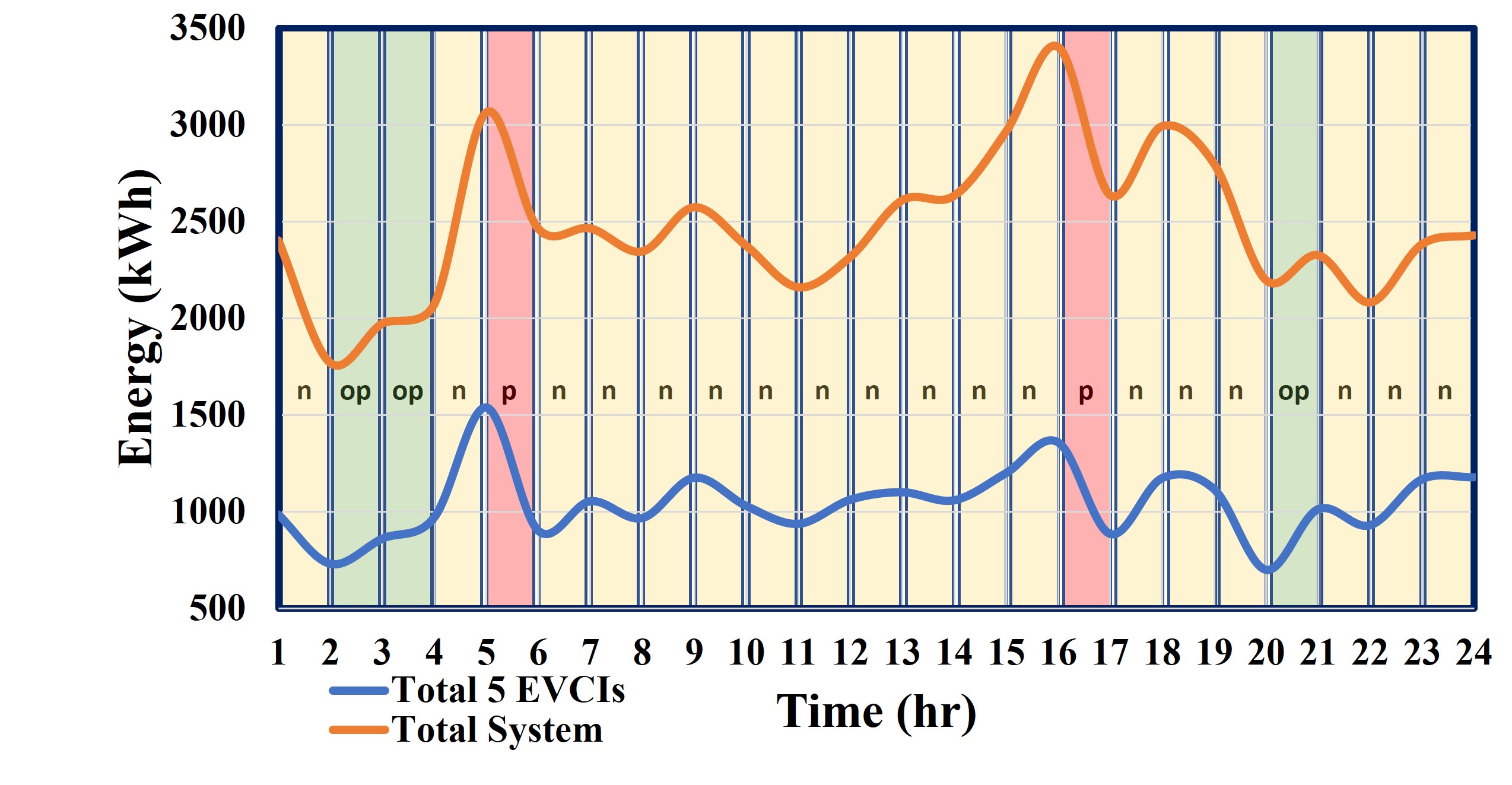}
  \caption{Total distribution system energy and total energy of all 5 EVCIs for Day 1 with three different time periods; normal hours (n) in yellow, of-peak hours (op) in green, peak hours (p) in red}
  \label{fig:system_energy}
\end{figure}

The combined energy of all 5 EVCIs and the total system energy is shown in Fig. \ref{fig:system_energy}. Based on the total EVCIs energy, the pricing of the EV charging is computed for peak $p$ hours, normal $n$ hours and off-peak $op$ hours represented in red, yellow, and green colors, respectively. Now we have the time-series data of the system parameters, such as voltage, power, power loss, and energy, we can perform the cost analysis of the EVCIs.
\subsection{Cost Analysis and Price Prediction Results}

\begin{figure}
  \centering
  \includegraphics[width=3.25in]{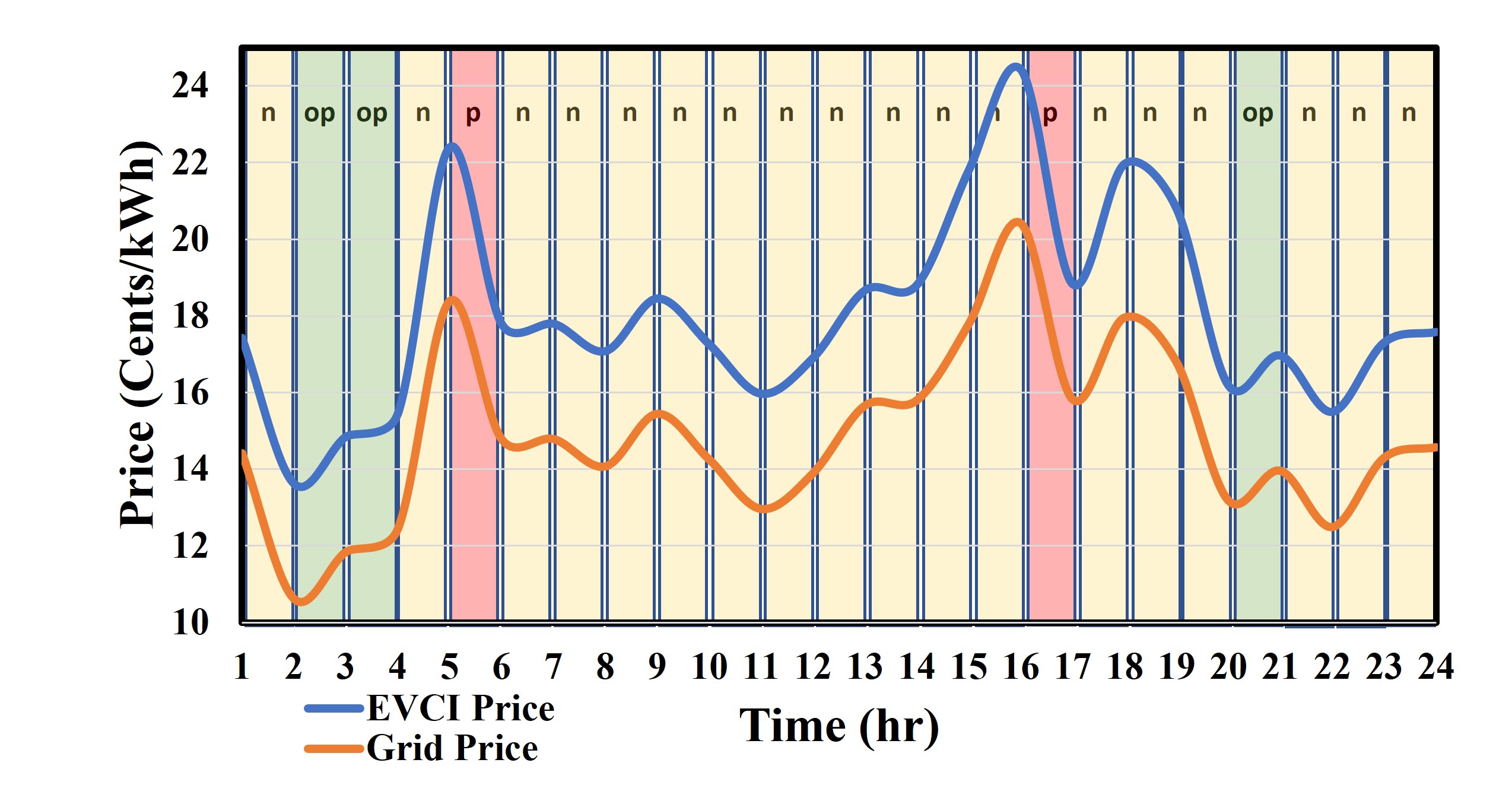}
  \caption{Hourly variations in grid price and EVCI price for Day 1 with three different time periods; normal hours (n) in yellow, of-peak hours (op) in green, peak hours (p) in red.}
  \label{fig:Price}
\end{figure}

\begin{figure}
  \centering
  \includegraphics[width=3.25in]{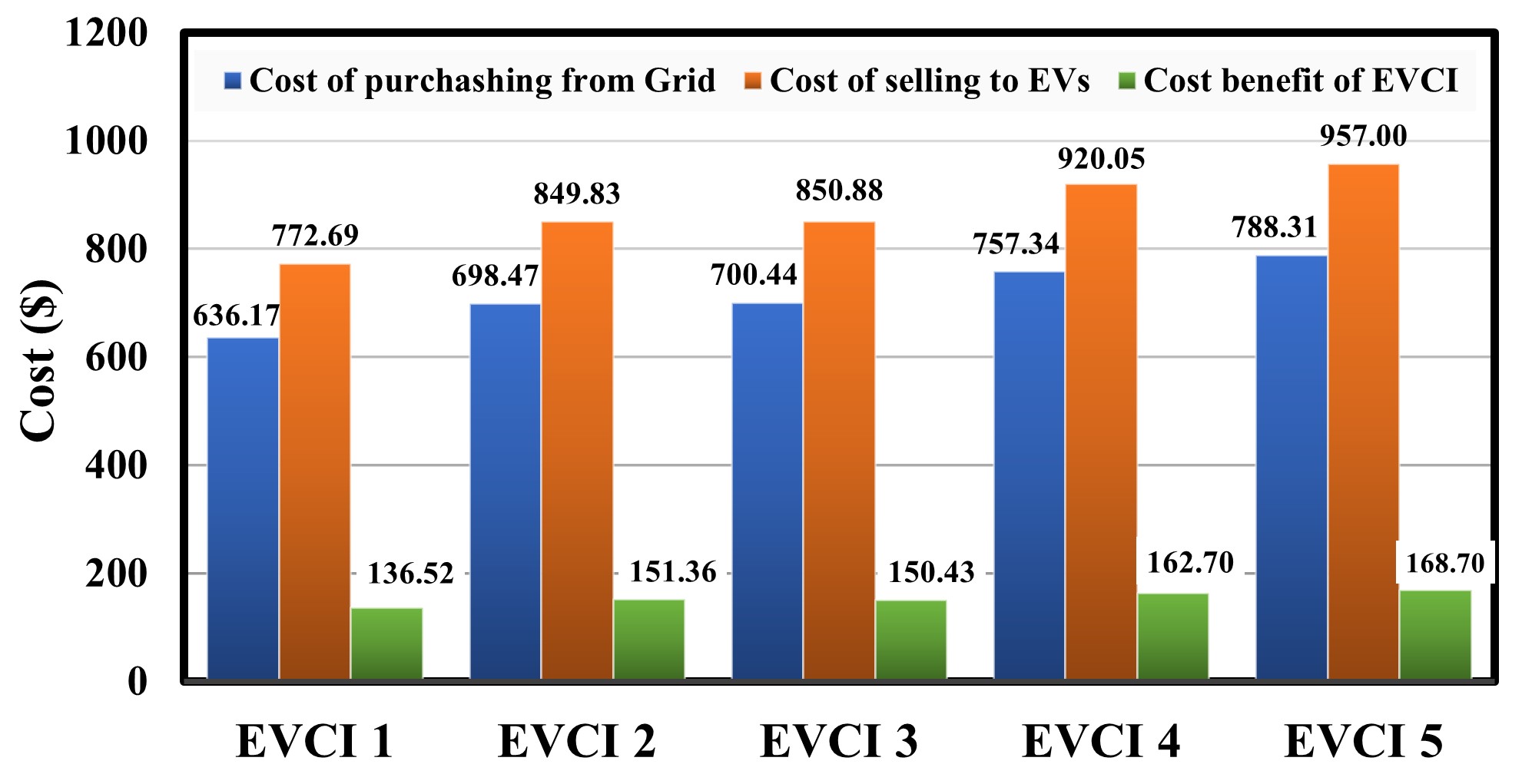}
  \caption{Comparison of the total cost of all 5 EVCIs for Day1 taking into consideration the cost of purchasing from the grid and cost of selling to EVs and the resulting cost benefits of EVCIs}
  \label{fig:EVCS_cost}
\end{figure}

We consider that the utility has set a real-time dynamic price that changes every hour. Now, the EVCI operator receives this information from the DS operator and accordingly sets the selling price at all five EVCIs. For simplicity, we consider one single price for all EVCIs by considering the total EV load at all 5 EVCIs; we can also compute the price for each EVCI, considering the EV load at each EVCI individually. Based on our pricing model, the prices for Day1 are considered as follows; fixed price for charging in the EV is 2\mbox{\textcentoldstyle}/kWh and the time-varying price is 2 to 8 \mbox{\textcentoldstyle}/kWh. These prices are considered based on the assumption of lower EV charging rates to promote EV adoption in the future. The EVCI operator has the flexibility of varying the time-of-use periods and their price, based on dynamic grid price and the EV traffic at the charging infrastructure. We observe the price variations of the grid and the EVCI in Fig. \ref{fig:Price}. Based on these prices, the EVCI sells the energy to the EVs and makes a profit. We can observe in Fig. \ref{fig:EVCS_cost}, that all EVCIs make a profit of \$136 to \$168 for Day 1.
\begin{figure}
  \centering
  \includegraphics[width=3.25in]{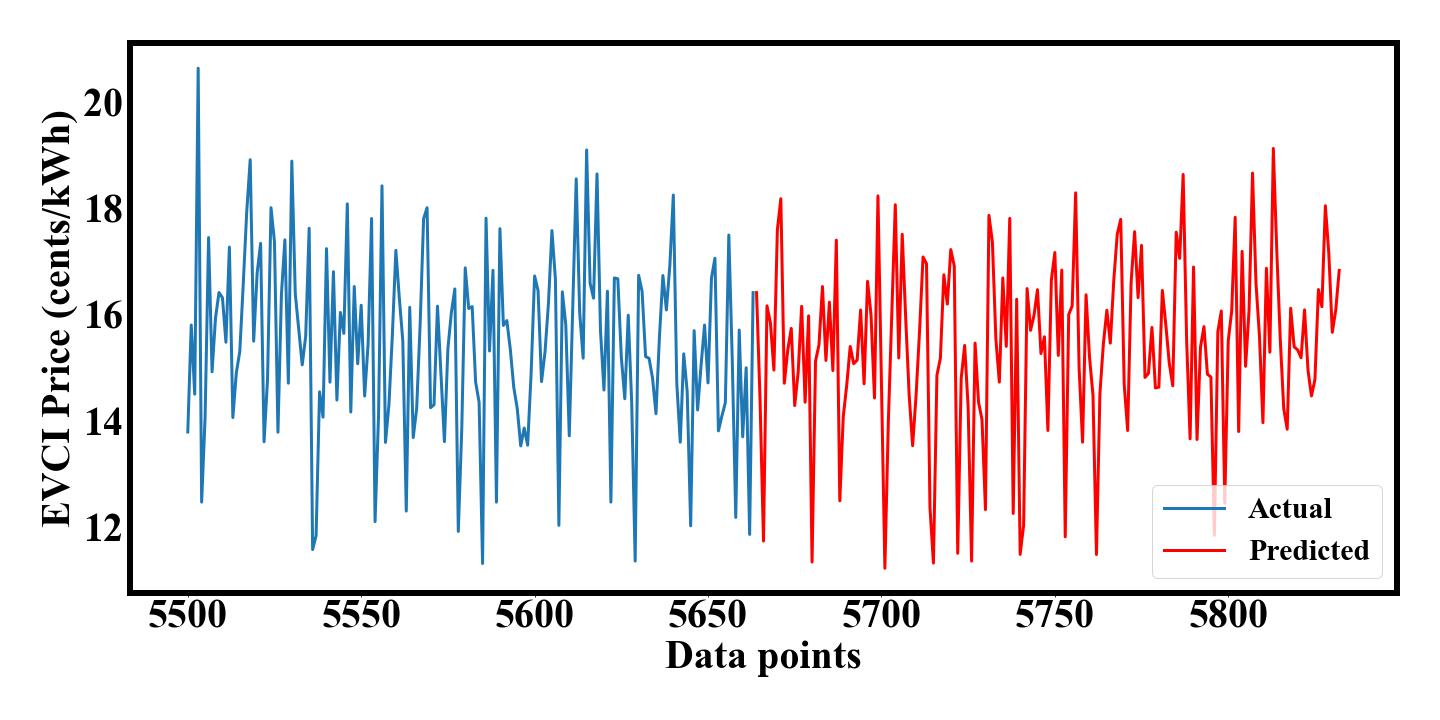}
  \caption{Cost prediction of EVCI price for 7 days (5665 to 5832 data points).}
  \label{fig:prediction_trained}
\end{figure}
\begin{figure}
  \centering
  \includegraphics[width=3.25in]{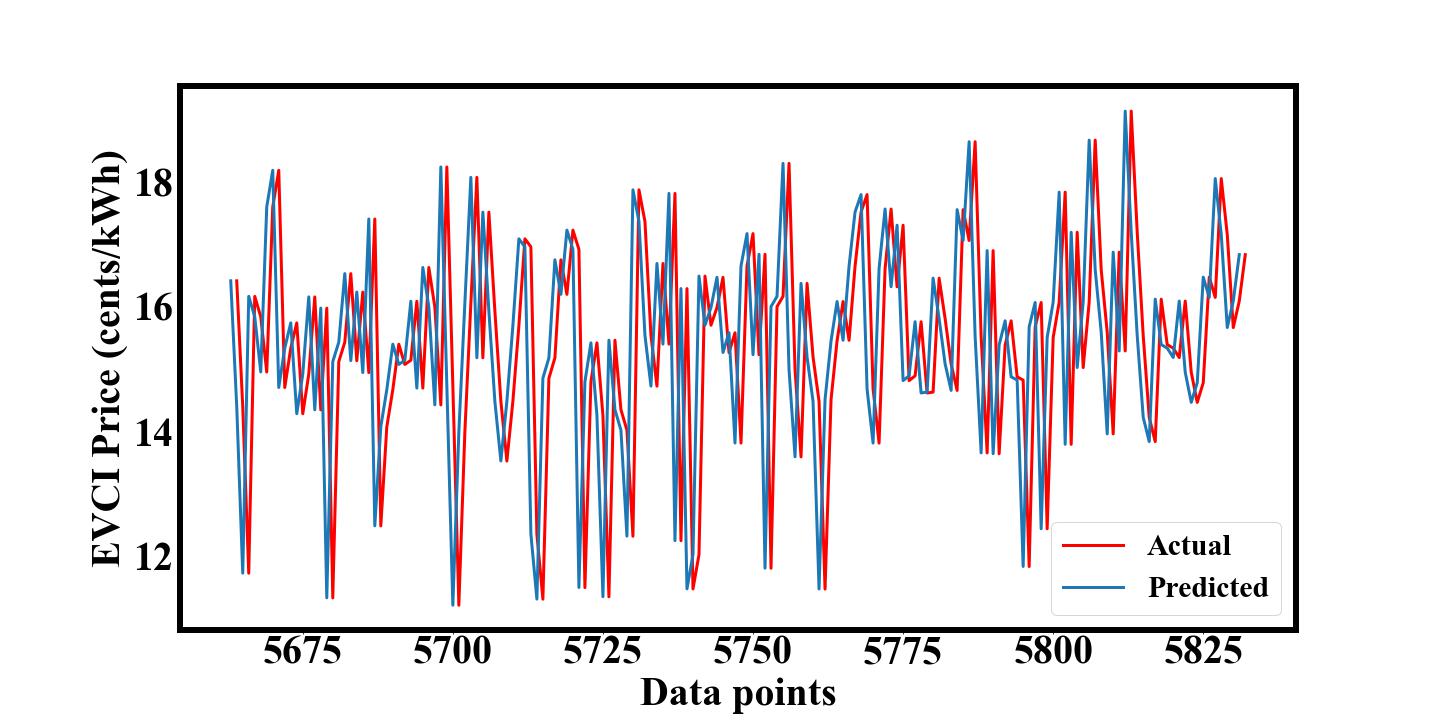}
  \caption{Comparison of the predicted data with known actual data for validation.}
  \label{fig:prediction_comp}
\end{figure}

As the time-varying data is for 8 months with 24 data points in one day, this gives a total of 5832 data points. For predicting the EVCI price, we perform the simulation to obtain the energy data of the system and compute the EVCI price for all 5832 data points. We use the first 5664 data points to obtain the parameters of ARIMA and use the remaining 168 data points, i.e., 7 days of data, to analyze the prediction. Using the 5664 data points, the parameters of $(p,q,d)$ are found to be $(0,2,1)$. We also checked the probability distribution of the residual errors and found it to be Gaussian distribution, which is an important check when using ARIMA model. The EVCI price for the last 7 days is shown in Fig. \ref{fig:prediction_trained}. We now compare this with the known values as shown in Fig. \ref{fig:prediction_comp}; the prediction is validated through the test scores given in Table. \ref{tab:Scores}; the R-squared value is 0.9999, which is close to 1 (maximum), indicating good prediction. We have also evaluated our model for the standard 7:3 ratio of the training to test set and found that the prediction scores are similar to that of the 5664:168 ratio. This implies the proposed ARIMA-based prediction model for the EVCI price could be utilized for predictions extending beyond one week.

\begin{table}[H]
\centering
\caption{Prediction Scores.}
\begin{tabular}{c c}
\toprule
Test & Score\\
\midrule
RMSE & 0.001000000000 \\
R\textsuperscript{2}  & 0.999999797385 \\
MAE  & 0.000663607216 \\
\bottomrule
\end{tabular}
  \label{tab:Scores}
\end{table}

\section{Conclusion}
\label{section:Conclusion}

We proposed an efficient strategy for the allocation of five fast-charging EVCIs in a distribution system by using a modified multi-objective PSO algorithm with minimum power loss and minimum voltage deviation. The proposed MOPSO algorithm demonstrates consistency in obtaining a global optimal solution and can be applied to a larger network. Furthermore, a time-series analysis of the distribution system under varying load and EV charging conditions demonstrates that the proposed method is effective in reducing power loss and voltage deviation. A cost model for the EVCI was developed considering the real-time dynamic pricing set by the utility. The prediction of EVCI price using ARIMA was proposed and validated. This can be used to assist the EVCI operator in forecasting cost benefits and offer discounts to EV customers at different time periods to increase utilization and profit. We plan to further investigate the system for various EV models and higher charger ratings and also perform a comprehensive comparative analysis of our proposed method with other popular optimization methods in future work.

Additionally, this work delves into the financial aspects of EVCIs under real-time pricing conditions. Utilizing an autoregressive integrated moving average (ARIMA) model to forecast dynamic prices, we found an impressively high R-squared value of 0.9999. This finding opens up possibilities for charging station operators to create promotional strategies to encourage utilization and boost profits.

For the simulation of data for EVs arriving at the charging station, a few assumptions were considered like a random number of EVs, random arrival times, and random initial charging states which do not affect the performance of the proposed methodology when evaluated for different conditions or data. Also, it is worth mentioning that with the availability of real-world data, we could obtain results that are more accurate and practical compared to simulated data. Thus, the proposed methodology was evaluated using simulated data and has not accounted for scenarios such as vehicle-to-grid and integration of distributed energy sources that will be considered for future works.

As a next step, researchers could investigate the implications of incorporating renewable energy sources, like solar or wind power, into the distribution system to enhance the performance and sustainability of EVCIs. Furthermore, examining the potential benefits of integrating energy storage systems may reveal new ways to optimize EV charging infrastructure under fluctuating demand and generation conditions. Overall, our research provides valuable insights into EV infrastructure planning and contributes to the development of a more sustainable and efficient transportation system.

\bibliographystyle{IEEEtran}
\bibliography{main}

\end{document}